\documentclass[fleqn,12pt]{wlscirep}
\usepackage[utf8]{inputenc}
\usepackage[T1]{fontenc}
\usepackage{amsfonts}

\usepackage[
    justification=justified
]{caption}

\title{Reconstructing commuters network using machine learning and urban indicators}

\author[1,*]{Gabriel Spadon}
\author[1]{Andre C. P. L. F. de Carvalho}
\author[1]{Jose F. Rodrigues-Jr}
\author[1,2,+]{Luiz G. A. Alves}
\affil[1]{University of Sao Paulo, Institute of Mathematics and Computer Sciences, Sao Carlos -- SP, 13566-590, Brazil}
\affil[2]{Northwestern University, Department of Chemical and Biological Engineering, Evanston -- IL, 60208-3112, USA}

\affil[*]{spadon@usp.br}
\affil[+]{lgaalves@northwestern.edu}


\begin{abstract}
Human mobility has a significant impact on several layers of society, from infrastructural planning and economics to the spread of diseases and crime. Representing the system as a complex network, in which nodes are assigned to regions ({\it e.g.}, a city) and links indicate the flow of people between two of them, physics-inspired models have been proposed to quantify the number of people migrating from one city to the other. Despite the advances made by these models, our ability to predict the number of commuters and reconstruct mobility networks remains limited. Here, we propose an alternative approach using machine learning and 22 urban indicators to predict the flow of people and reconstruct the intercity commuters network. Our results reveal that predictions based on machine learning algorithms and urban indicators can reconstruct the commuters network with $90.4\%$ of accuracy and describe $77.6\%$ of the variance observed in the flow of people between cities. We also identify essential features to recover the network structure and the urban indicators mostly related to commuting patterns. As previously reported, distance plays a significant role in commuting, but other indicators, such as Gross Domestic Product (GDP) and unemployment rate, are also driven-forces for people to commute. We believe that our results shed new lights on the modeling of migration and reinforce the role of urban indicators on commuting patterns. Also, because link-prediction and network reconstruction are still open challenges in network science, our results have implications in other areas, like economics, social sciences, and biology, where node attributes can give us information about the existence of links connecting entities in the network.
\end{abstract}
\begin{document}

\flushbottom
\maketitle

\thispagestyle{empty}
\newpage

\section*{Introduction}

Humans move daily to work, do business, have leisure, meet people, and perform routine activities. Modeling human mobility is vital to better allocate resources and to improve the impacts of human activities in the community (nearby people) and the environment (cities and nature). From physics and mathematics to geography and social sciences, several researchers have tried different approaches to understand the impacts of human movement on society and vice-versa~\cite{Barbosa2018:HumanMobility,Ullman1980:Geography}. Human mobility is shaped by the urban organization~\cite{Bettencourt2013:ScalingCities} and can also change cities as an effect of traffic congestion~\cite{Louf2013:ModelingPolycentric}. Mobility patterns are associated with energy use~\cite{Trenchard2016:BiologicalSystems, Helbing2015:ComplexityScience}, the spread of diseases~\cite{Eubank2004:DiseaseOutbreaks, Colizza2006:Epidemics, Balcan2009:InfectiousDiseases, Yang2018:EpidemicSpreading}, the occurrence of crimes~\cite{Caminha2017:CrimeProxy,Spadon2017:UrbanCrimes}, and others. Therefore, good predictive models can, for instance, help improve daily human activities with better urban planning, and also help policy-makers with more informed decisions to intervene in the disease spreading and crime.

Prediction of migration from one area to another, at different time scales, is one of the most critical challenges in human mobility. It is common to represent the system as a spatial complex network, where each node represents a region ({\it e.g.}, city) and the links between two regions indicate the flow of people. Thus, physics-inspired models such as the gravitation~\cite{Zipf1946:IntercityMovement,Jung2008:GravityModel} and radiation models~\cite{Simini2012:UniversalModel,Masucci2013:GravityRadiation} have been used to predict the edges of the network as well as their weights (the number of people migrating). These models assume that the number of people going from one region/node to another decays with the distance separating them and is proportional to the populations' masses of these regions~\cite{Zipf1946:IntercityMovement,Jung2008:GravityModel,Simini2012:UniversalModel,Masucci2013:GravityRadiation}. However, this assumption often fails to accurately describe the flow of people because of other factors that can increase or decrease mobility, such as the underlying transportation network~\cite{Ren2014:CommuterFlows}, socioeconomic aspects~\cite{Ravenstein1885:MigrationLaw,Stouffer1940:InterveningOpportunities,Bettencourt2013:ScalingCities,Louail2015:SpatialStructure}, and transport congestion~\cite{Louf2013:ModelingPolycentric}. Usually, these models are limited to predicting the weights of existing links, and they overestimate the number of connections between nodes when dealing with sparse mobility networks. Therefore, these limitations make it hard to generalize the model on unseen data and to reconstruct the structure of human mobility networks. 

Link prediction and network reconstruction is a very active area of research in network science~\cite{Lu2011:LinkPrediction}.
Most of the literature on link prediction is based on evaluating the similarity between nodes and suggesting missing links. For instance, the metrics used to evaluate the existence of links include, but are not limited to, the number of common neighbors~\cite{newman2001clustering,kossinets2006effects}, the existence of short paths between nodes~\cite{jeh2002simrank}, measures of centrality~\cite{Fu2018:LinkWeightPrediction}, and hierarchical structures within the network~\cite{clauset2008hierarchical}. Reconstruction techniques also include methods based on maximum-entropy distribution regarding the node degree and the flow strength as constraints~\cite{mastrandrea2014enhanced,squartini2015unbiased} and Bayesian inference of links based on edge data information~\cite{guimera2009missing, peixoto2018reconstructing}. In general terms, these methods use network-based metrics to infer the existence (or non-existence) of links, which significantly differs from predictive models based on meta-data attributes such as the population size or the distance between nodes. 

A few models can be found in the literature where the node's attributes are used as complementary information to predict links. For instance, in the context of social contact networks, mobility patterns such as co-location of individuals was used to evaluate the probability of individuals to be connected in a social network~\cite{Wang2011:HumanMobility}, restaurant reviews were used to predict links in a taste similarity network~\cite{Xuan2018:CuisineChoices}, and the frequency of programming code commits was used to predict the mobility in cyberspace of projects~\cite{Xuan2014:CallGraphs}. The focus of our work is in the class of methods that can use node's attributes, such as urban metrics, as input data to fit models that can predict links and can be further extended to other data sets where the dependent variable is unknown. 

Recently, an unprecedented amount of data related to human behavior and cities became available, from GPS tracking of mobile phones to census data of thousands of people~\cite{Makse1995:UrbanGrowth, Jung2008:GravityModel, Thiemann2010:Borders, Roth2011:UrbanMovements, Barthelemy2011:SpatialNetworks}. These data promoted intense research on human mobility~\cite{Barbosa2018:HumanMobility}, including transportation networks~\cite{Guimera2005:AirTransportation}, commuters networks~\cite{Ren2014:CommuterFlows}, and network models of migration~\cite{Lee2014:Matchmaker}. On the other hand, urban indicators were used to describe scaling in cities~\cite{Bettencourt2013:ScalingCities}, to measure the performance of cities~\cite{Alves2015:CityPerformance} and the similarity among different municipalities~\cite{Domingues2017:Topological, Spadon2018:TopologicalCharacterization}, and to describe crime-related phenomena~\cite{Alves2013:CrimeScaling, Bettencourt2013:ScalingCities, Caminha2017:CrimeProxy}. However, a connection between human mobility and urban indicators, such as unemployment rate and Gross Domestic Product (GDP) is still missing. Understanding the influence of these indicators on the individual choices on daily commuting to work could help us to predict the flow of people between different areas and reconstruct the structure of the commuters network. Such a gap has led us to frame the question we want to answer: {\it how to quantify the number of people commuting in between cities taking into account a more comprehensive set of indicators, beyond distance and population, that might attract or repel commuters?}

In this study, we propose an alternative approach to deal with the reconstruction of commuters networks using supervised machine learning to understand the relationships between urban indicators and the structure of commuters networks. We perform an analysis based on 22 urban indicators of $5,565$ Brazilian municipalities, together with the daily number of people commuting between every city in the data set. We show that the gravitation and radiation models have limited predictive power to classify whether there is a link between two cities or not and to quantify the flow (number of people commuting) between cities. In contrast, we show that predictions based on machine learning algorithms and urban indicators can reconstruct the commuters network with $90.4\%$ of accuracy and describe $77.6\%$ of the variance observed in the flow of people between cities. Further, we interpret the machine learning results using SHapley Additive exPlanations (SHAP) values~\cite{Lundberg2017:SHAP} to quantify the importance of the urban indicators in predicting human mobility. We show that distance is a critical metric for predicting human mobility, but other indicators, like GDP and unemployment rate, also play a significant role in attracting/repelling people to a particular area. Our approach provides a better way to quantify human mobility and shed new lights on the role of urban indicators on commuting patterns. Because link-prediction and network reconstruction are still open challenges in network science, our results have implications in other areas, like economics, social sciences, and biology, where node attributes can give us information about the existence of links connecting network's entities.

\begin{figure}[!b]
    \centering
    \includegraphics[width=0.9\textwidth]{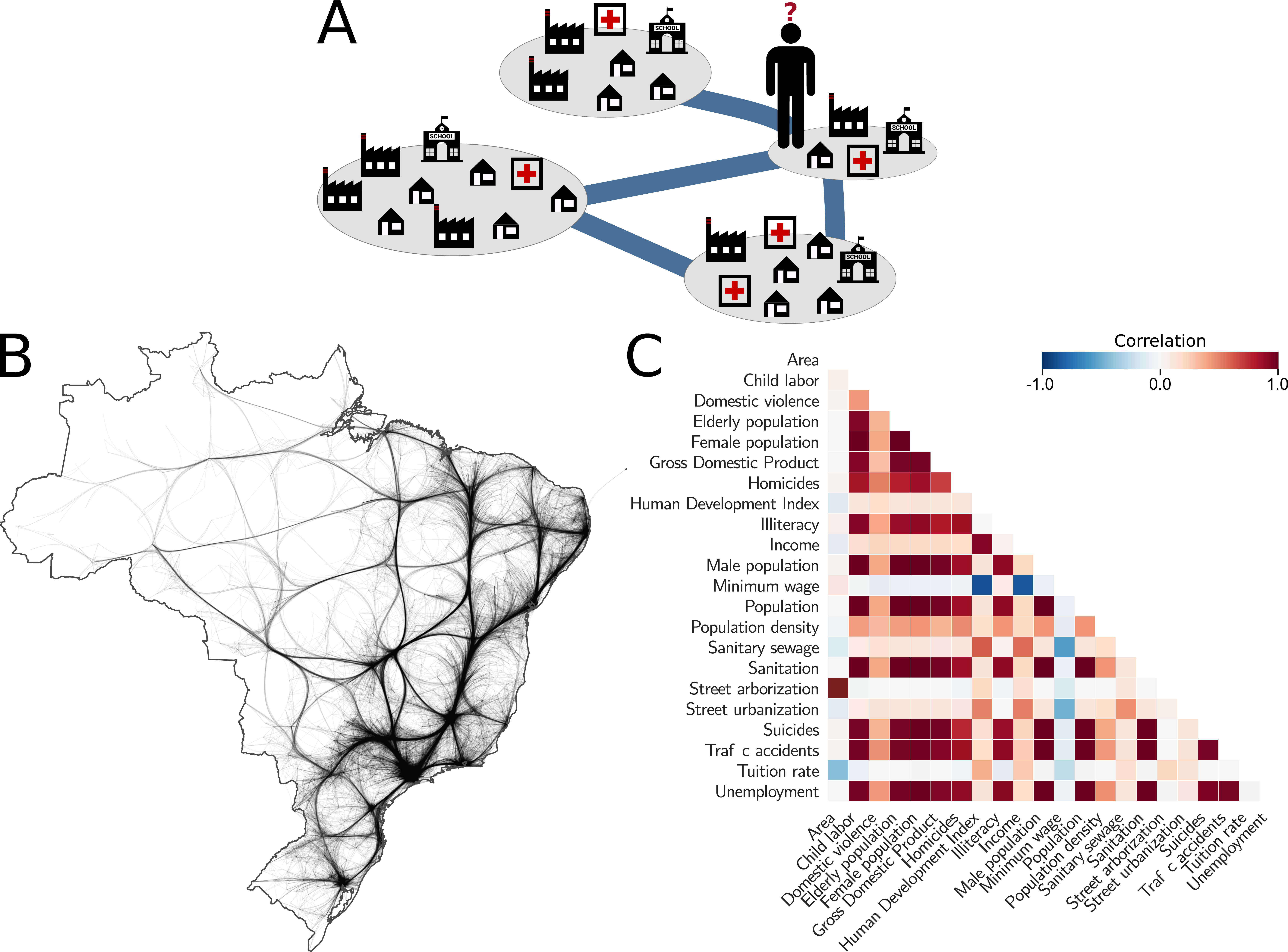}
    \caption{Commuters network and urban indicators related to human mobility. {\bf (A)} Illustration of the work-mobility problem when people have to choose where to work based on distance, job opportunities, housing prices, and other variables related to urban systems. {\bf (B)} Brazilian commuters network illustrated through a kernel-based Edge Bundling technique~\cite{Moura2015:3DHEB} where the thickness of the edges represents the flow intensity. {\bf (C)} Correlation analysis of Brazilian urban indicators considering those that describe the population, labor, tuition rate, economy, sanitation, infrastructure, violence, and others. Reddish colors correspond to positively correlated indicators and bluish colors to negatively correlated ones.}%
    \label{fig:1}
\end{figure}

\section*{Results}

We started our study from the observation that people move from one city $s$ to work in another city $t$, taking into consideration the advantages of living nearby or far away from where they work. In Fig.~\ref{fig:1}A each node represents a city that offers different opportunities (in terms of jobs), costs of living (number of houses available), and other factors, such as hospitals, schools, and urban parks, to name a few. Two given cities, $s$ and $t$, are separated by a distance $r_{st}$, and people have a cost to commute from their home city to the city where they work. Existing migration models assume that the flow of people from city $s$ to city $t$ is proportional to their populations and decays with the distance. However, plenty of other factors play an important role when deciding where to live or work.

To quantify the number of people commuting in between cities taking into account a more comprehensive set of indicators, beyond distance and population, we collected data about the number of people commuting from one city to another (pendular migration), in 2010, considering all the $5,565$ Brazilian cities, together with 22 urban indicators related to these municipalities. This data is provided and maintained by the Brazilian Institute of Geography and Statistics (IBGE)~\cite{IBGE}. The data consists of the number of people that are living in a city $s$ and commuting to work in a city $t$ daily, and of urban indicators describing the cities in terms of population, labor, tuition rate, economy, sanitation, infrastructure, and violence, to name a few. We suppose that these indicators can be associated with a high or low number of commuters. Thus, less economically developed cities, usually those with fewer job opportunities, inferior infrastructure, and possibly higher indices of violence, would attract a lower number of people. On the other hand, those with a higher income, larger population, better social development, and infrastructure, might offer better job opportunities and would attract a higher number of people.

To investigate the flow of people commuting from one city to another, we assigned a node for each city and, for each non-zero flow, an edge with weight $w_{st}$ equal to the number of people commuting from city $s$ to $t$ in the year 2010. Fig.~\ref{fig:1}B illustrates the Brazilian commuters network using a kernel-based edge bundling technique~\cite{Moura2015:3DHEB}. To illustrate our urban indicators, in Fig.~\ref{fig:1}C, we present the Pearson correlation coefficient between the 22 urban indicators.

\subsection*{State-of-the-art models}
\label{subss:review}

Next, we investigate the limitations of two models widely used to describe the flow of people between cities, namely, the gravitation model and the radiation model.

\noindent\textbf{Gravitation model.} Inspired by Newton's law of universal gravitation, it has been first used to predict the trading between nations; the model is defined as:
\begin{equation}
    T_{st}^{GM} = C \frac{m_{s}^{\alpha} n_{t}^{\beta}}{r_{st}^{\gamma}},
    \label{eq:gravitation-model}
\end{equation}
\noindent where $m_{s}$ and $n_{t}$ are the supply and demand forces of nations $s$ and $t$, respectively~\cite{Tinbergen1962:WorldEconomy}. This model was widely applied to predict population movement~\cite{Jung2008:GravityModel}, economic relations between countries~\cite{Helpman2008:TradeFlows}, cargo shipping volume~\cite{Kaluza2010:CargoShip}, and long-distance phone calls~\cite{Expert2011SpatialNetworks}. In the context of migration, $T_{st}^{GM}$ is the number of people commuting, $s$ and $t$ are cities, $m_{s}$ and $n_{t}$ are the population masses of these cities, and $r_{st}$ is the distance between them. Thus, the interaction between two municipalities is directly proportional to the population masses and inversely proportional to the distance. Such a model depends on the parameters $C$, $\alpha$, $\beta$, and $\gamma$, which can be estimated through Ordinary Least Squares (OLS) applying the logarithm operation on both sides of the formula. Although the gravitation model provides a good fit for a wide variety of scenarios, there are still some unsolved problems. For instance, Simini {\it et al.}~\cite{Simini2012:UniversalModel} pointed out that:
\begin{itemize}
    \setlength\itemsep{0em}
    \item The model's equation lacks a rigorous mathematical derivation;
    \item The augmentation of the model is unrestricted, what would scale the number of parameters;
    \item The model fails when the data is not sufficient to estimate its parameters;
    \item The flow is only a function of population and distance, not including any other characteristic inherent to the cities and their neighbors;
    \item The number of commuters increases without limit as the population of the target city grows, becoming higher than the total population of the source city; and,
    \item The model is deterministic, being unable to estimate fluctuations in the number of people commuting.
\end{itemize}

\noindent\textbf{Radiation model.} Formulated in terms of the radiation and absorption processes from physics, this model was proposed to solve the problems inherent to the gravitation model~\cite{Simini2012:UniversalModel}. It is based on a diffusion process in which an object in a specific location emits particles, having nearby objects with a different probability of absorbing these particles, which varies according to the forces acting upon them. In the context of commuters, the objects are the cities, and the particles are the people commuting. Thus, cities are radiating commuters, which are absorbed by neighboring cities.

In contrast to the gravitation model, in the radiation model, the distance between cities is not the major limitation to commuters; differently, the limitation arises from the supply and demand of commuters concerning the neighboring cities. The idea of supply and demand comes from the theory of intervening opportunities; this theory defines that mobility patterns are more influenced by the commuters' opportunity to establish in the target city than the distance to such a city~\cite{Stouffer1940:InterveningOpportunities}. The approximate number of intervening opportunities $p_{st}$ is calculated according to the population of the source $s$ and target $t$ cities. It also takes into account the population of all the neighbors of the source city in a maximum radius of up to the distance to the target city. Thus, the radiation model can be defined as:
\begin{equation}
    T_{st}^{Rad} = T_{s} \frac{m_{s} n_{t}}{(m_{s} + p_{st})(m_{s} + n_{t} + p_{st})},
    \label{eq:radiation-model}
\end{equation}
\noindent where $m_{s}$ is the population of the source city, $n_{t}$ the population of the target city, and $p_{st}$ is the total population in the circle of radius $r_{st}$ centered at city $s$.

The only parameter of the radiation model, $T_{s}$, is the scaling factor of the outgoing number of commuters from the source city. The value of $T_{s}$ can be estimated by adjusting the equation $T_{s} = \alpha m_{s}$ to the data using OLS, in which $m_{s}$ is the population of the source city and $\alpha$ is the intersection point of the fitting between the population size and the number of commuters from every city in the whole data set. The major limitation of the model is that $T_{s}$ depends on the number of commuters. Thus, the model requires the correct information (or an estimation) about the number of people commuting between cities, which makes it difficult to generalize the model to unseen data.

Similarly, Ren {\it et al.}~\cite{Ren2014:CommuterFlows} proposed the cost-based radiation model to predict the flow of commuters in spatial networks. Although the model has been reported as more efficient than its predecessor, its predictive ability is still limited to cases where there is information about the existence of links. As a consequence, the radiation model and its cost-based version are not able to reconstruct the structure of the commuters network.

It is also worth to mention that most of these models (gravitation and radiation models) were validated through correlation analysis, rather than using the variance between the predicted and real values; this fact causes an overestimation of the capability of the model in predicting the flow of people between cities. The variance is more error-sensitive than the correlation coefficient ({\it e.g.}, Pearson, Spearman, or Kendall). Moreover, when assessing the results, some authors take into account only existing edges to compare with the predictions, causing, once more, an optimistic estimation of their predictive performance.

Next, to further confirm our claims, we applied the gravitation and radiation models to predict the number of people in the Brazilian commuters network. Our first observation is that these models (Eqs.~\ref{eq:gravitation-model} and~\ref{eq:radiation-model}) are not able to reconstruct the unweighted projection of the commuters network structure, because any pair of cities with non-zero population would have an edge, resulting in a fully connected network. If one considers faraway cities, it is possible to have flow close to zero, but for typical distances between Brazilian cities (average value of $\sim 1,000 km$), we would still find a non-zero flow. The resulting networks generated by these models are far from the sparse networks observed in real data. Thus, we quantified the number of commuters considering only edges that we know beforehand to have a non-zero flow. We considered the distance to be the length of the geodesic path (in kilometers) between two given cities, which is calculated from their coordinates provided by the Brazilian census data~\cite{IBGE}.

Our evaluation started by fitting the models (Eqs.~\ref{eq:gravitation-model} and~\ref{eq:radiation-model}) to the data through OLS to estimate their descriptive parameters. Next, we evaluated the results using $R^2$-score (coefficient of determination) and Pearson correlation coefficient. The $R^2$-score measures the variance of a dependent variable that was predicted using independent variables; the metric is defined in the range~$]-\infty,~1]$. An $R^2$-score closer to $1$, means that the variance of the prediction concerning the real value is low~\cite{Carpenter1960:Statistics}. Notice that this metric can be negative since the model can be arbitrarily wrong. It is worth mentioning that, along with the text, we express the $R^2$-score through percentage ({\it e.g.}, 14\% instead of 0.14) because we use it to discuss the proportion of variance explained by the correlation between the predicted and observed flow. The Pearson correlation coefficient $\rho$, in turn, measures the linearity of two variables regarding each other. The metric is defined in the range $[-1, 1]$, in which $-1$ indicates a perfect negative correlation, $0$ indicates no linear correlation and $1$ indicates a perfect positive correlation~\cite{Chiang2003:StatisticalAnalysis}. With very wrong predictions, one could find results that are very correlated with the empirical values ($\rho \approx 1$), misleading the evaluation of the predictive model performance.

\begin{figure}[!htb]
    \centering
    \includegraphics[width=0.5\columnwidth]{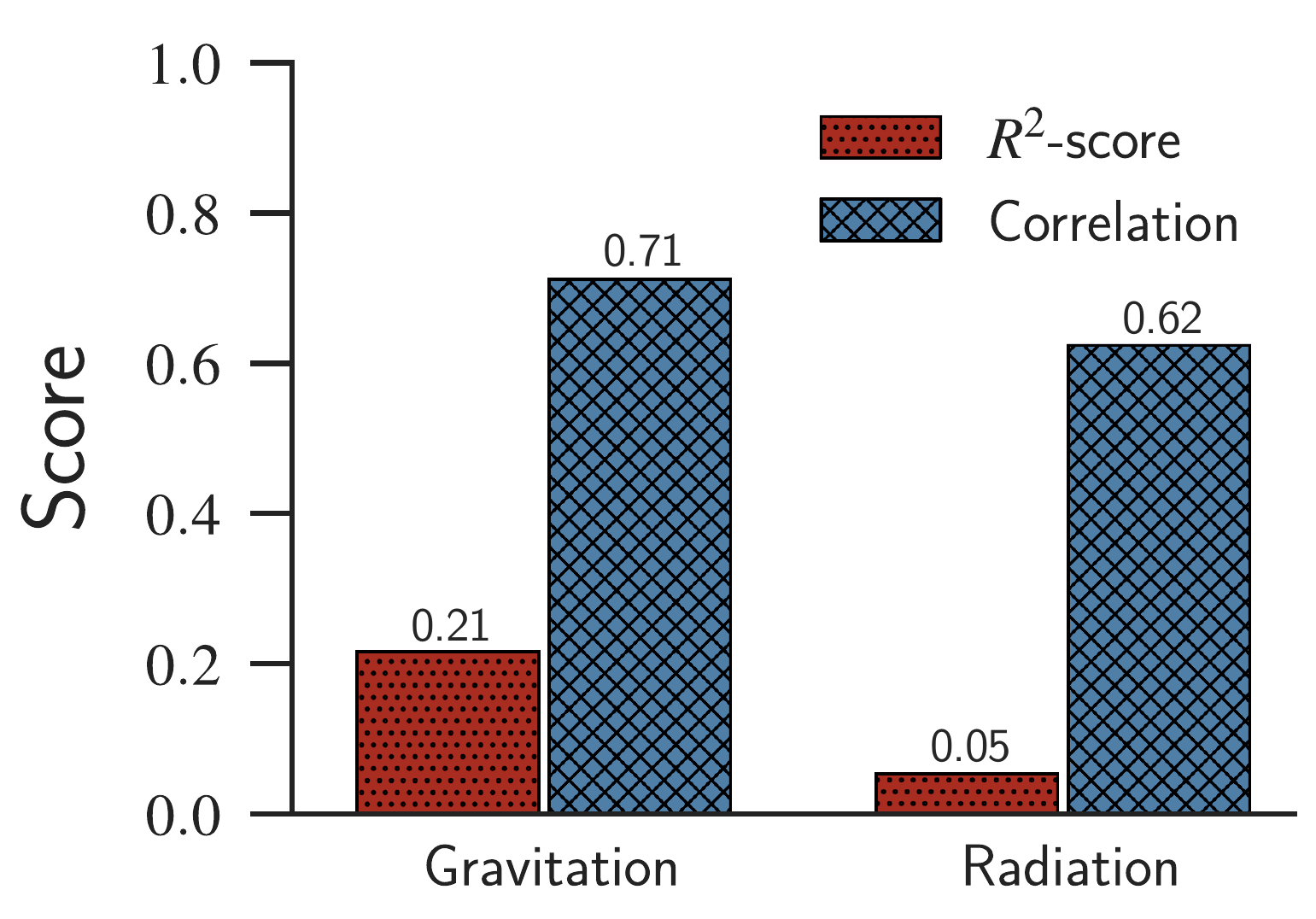}
    \caption{Evaluating the predictive ability of the gravitation and radiation models. We used two metrics, the coefficient of determination (referred to as $R^2$-score), and the Pearson correlation coefficient; both metrics consider the predicted flow and the observed flow of the models. Although the results show a linear correlation between the predicted and observed flows, the $R^2$-score reveals that the predicted values are still very noisy when compared to the real ones, suggesting that neither model is accurate enough in reconstructing the commuters network.}
    \label{fig:2}
\end{figure}

Fig.~\ref{fig:2} presents the results related to the fitting of the gravitation and radiation models to our data. From this figure, we verify that the values predicted by both models are positively correlated (values greater than 0.60) with the real data, but the $R^2$-score is meaningless regardless of the model (values below 0.25). Despite the good correlation, the models were not able to correctly predict the data, and the values differ sharply from the real ones. The same conclusion was brought by Masucci {\it et al.}~\cite{Masucci2013:GravityRadiation}, who analyzed the ability of generalization and universality of both models with a data set of commuters in the surroundings of London, UK. In agreement with what we observed when evaluating the model using the Brazilian data set, neither model could satisfactorily fit their data. Thus, we confirmed that the gravitation and radiation models could not correctly describe the mobility pattern of the intercity commuters network.

\section*{Alternative modeling using machine learning}

Next, we propose the use of predictive models induced by machine learning algorithms to predict the number of intercity commuters and to reconstruct the structure of the corresponding commuters network. Our approach differs from the regular link prediction because it does not take into account any network-based topology metric as an independent variable. Instead, we reconstruct the network based on the distance $r$ between two cities, their populations' size and a set of urban indicators related to each city. To do so, we investigate two predictive tasks:
\begin{itemize}
    \setlength\itemsep{0em}
    \item Classification: induce a binary classifier able to predict whether a link between a pair of cities exists or not;
    \item Regression: induce a regressor able to predict the number of commuters between a pair of cities.
\end{itemize}

\subsection*{Reconstructing the commuters network structure using machine-learning classification}%

In our context, a binary classifier predicts whether a link between a city $s$ and a city $t$ exists or not based on the distance $r_{st}$, populations sizes $m_s$ and $n_t$, and more 21 urban indicators from each city, $U_i=\{u_{i_0}, \ldots, u_{i_{20}}\}$, $i \in \{s,t\}$, and $u_{i_j} \in {\rm I\!R}$ for $0 \leq j \leq 20$, resulting in a total of 45 urban indicators (referred to as features).
That is, given an ordered pair $\left<city_s, city_t\right>$ whose urban indicators define a set $S_{st}=\{r_{st}, m_s, n_t, U_s, U_t\}$, $S_{st} \in {\rm I\!R}^{|S|}$, a binary classifier refers to a function
\begin{equation}
    Class:{{\rm I\!R}^{|S|}} \to \{0,1\}
\end{equation}
\noindent in which, $0$ indicates no link between $s$ and $t$, and $1$ indicates the opposite.

To find the best classifier, we first employed the holdout approach, splitting the data into $70\%$ for training and $30\%$ for testing. Then, we sampled the training data using stratified $k$-fold cross-validation, with $k=5$. The training data was used in the model selection, feature selection, and hyperparameter tuning. The test set was used only in the final step, after the hyperparameter tuning, to evaluate the predictive performance, generalization, and universality of the model in an unseen data set.

Subsequently, we tested 34 classification algorithms with default hyperparameter values from the {scikit-learn}~\cite{scikit-learn} and eXtreme Gradient Boosting ({XGBoost})~\cite{Chen2016:XGBoost} libraries, from which only 27 were able to fit the data and provide the resulting accuracy. The 7 removed algorithms (see Supplementary Table~1) fail to fit the data ({\it i.e.}, to converge) without a pre-tuning of hyperparameters, which is intentionally not covered by our methodology. We do not perform the pre-tuning of hyperparameter because it would exponentially increase the processing time of the experiments' pipeline with no improvement guaranteed.

\begin{figure}[!htb]
    \centering
    \includegraphics[width=0.9\columnwidth]{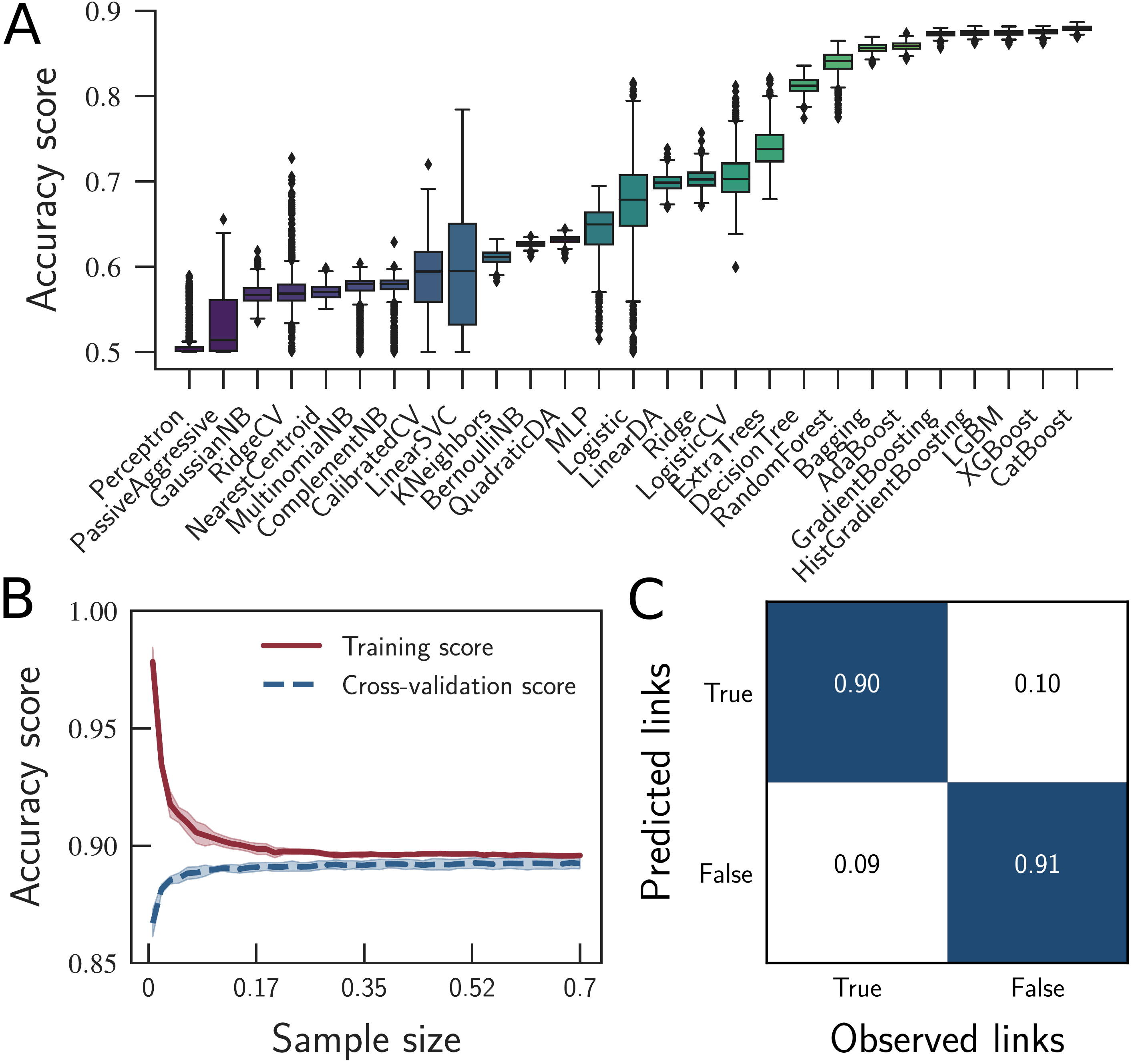}
    \caption{Performance of the classification algorithms in reconstructing the unweighted projection of the commuters network; see Supplementary Table~1 for a list of the classifiers' acronyms. {\bf (A)} The accuracy of the classification algorithms in determining whether a link between a given pair of cities exists or not. {\bf (B)} XGBoost learning curves varying the sample size. {\bf (C)} XGBoost confusion matrix after hyperparameter tuning, showing an overall accuracy of $90.4\%$. Specifically, the results showed $90\%$ of true positives, $10\%$ of false negatives, $9\%$ of false positives, and $91\%$ of true negatives.}%
    \label{fig:3}
\end{figure}

Following, to select the best among the 27 remaining classifiers, we used bootstrapping sampling to evaluate their predictive performance following an accuracy-based perspective. This procedure consists in feeding the model with randomly selected samples to assess its variance. Fig.~\ref{fig:3}A shows the results of the classifiers that passed our first test. The algorithms were sorted in ascending order according to their predictive accuracy. It is worth to mention that the best algorithms (rightmost) are the CatBoost, XGBoost, and Light Gradient Boosting Machine (LGBM), and the ones with the worst performance are the Perceptron, Passive Aggressive Classifier, and Gaussian Naive Bayes (NB). The average accuracy score in the experiment varies from 50.2\% in the worst case to 87.9\% in the best case --- see Supplementary Table 3. Notice that, the algorithm with the highest median score, lowest variance, and fewer outliers is the CatBoost. However, the difference between the CatBoost (first-placed) and XGBoost (second-placed) is irrelevant (see Supplementary Table~3), and the XGBoost is around fifty times faster per train iteration than the CatBoost. For this reason, we have chosen the XGBoost as the best algorithm for this problem.

Following, we evaluated the learning curve considering only the best classifier ({\it i.e.}, XGBoost), which showed an average accuracy score of 87.5\% on the training set, with little variance and no significant accuracy outliers. The learning curve assesses the model predictability by varying the size of the training set. The curve in Fig.~\ref{fig:3}B shows that 89\% of the training set is the right amount of data required to train the classifier and a classifier trained up to such amount of data yields an accuracy of about 89.2\%.

To reduce the computational cost in the induction of the predictive model using XGBoost, we used a threshold-based feature selection (see Methods), measuring the accuracy of the model by progressively removing features. We considered the degree of importance of a feature as the number of times the feature is used to split a node into two trees in the XGBoost algorithm. By doing so, we were able to remove 24 predictive features without reducing the model's predictive accuracy (see Supplementary Table 5).


Finally, we used the test set (the remaining 30\% of the data previously not used) to test the induced model. We first calculated the predictive accuracy of our classifier on the test data using the model with no grid-search optimization, finding $89.3\%$ of accuracy. The grid-search was able to improve this value by $1.1\%$ (see Methods), which, in a scenario with thousands of cities with millions of possibilities of commuters flowing between them, means more hundreds of thousands of correctly predicted links. Therefore, such an improvement resulted in an overall accuracy of $90.4\%$, as shown in the confusion matrix in Fig.~\ref{fig:3}C.

\subsection*{Reconstructing the weighted commuters network using machine-learning regression}

Deeper in the analytical scenario, we are interested not only in knowing whether a link exists or not but also in the weight $w_{st}$ of each link, which represents the flow of people from city $s$ to $t$. That is, given an edge $\left<city_s, city_t\right>$ and its corresponding set of indicators $S_{st}=\{r_{st}, m_s, n_t, U_s, U_t\}$, $S_{st} \in {\rm I\!R}^{|S|}$, we seek a function
\begin{equation}
    Weigh:{{\rm I\!R}^{|S|}} \to {\rm I\!N}
\end{equation}
\noindent where $Weight$ predicts the number of commuters between cities $s$ and $t$.

Similarly to the prediction of links, we first used holdout to split the data into $70\%$ for training and $30\%$ for testing. Subsequently, the training set was used for $5$-fold cross-validation, model selection, feature selection, and hyperparameter tuning, and the test set to evaluate our model in an unseen data set.

\begin{figure}[!htb]
    \centering
    \includegraphics[width=0.9\linewidth]{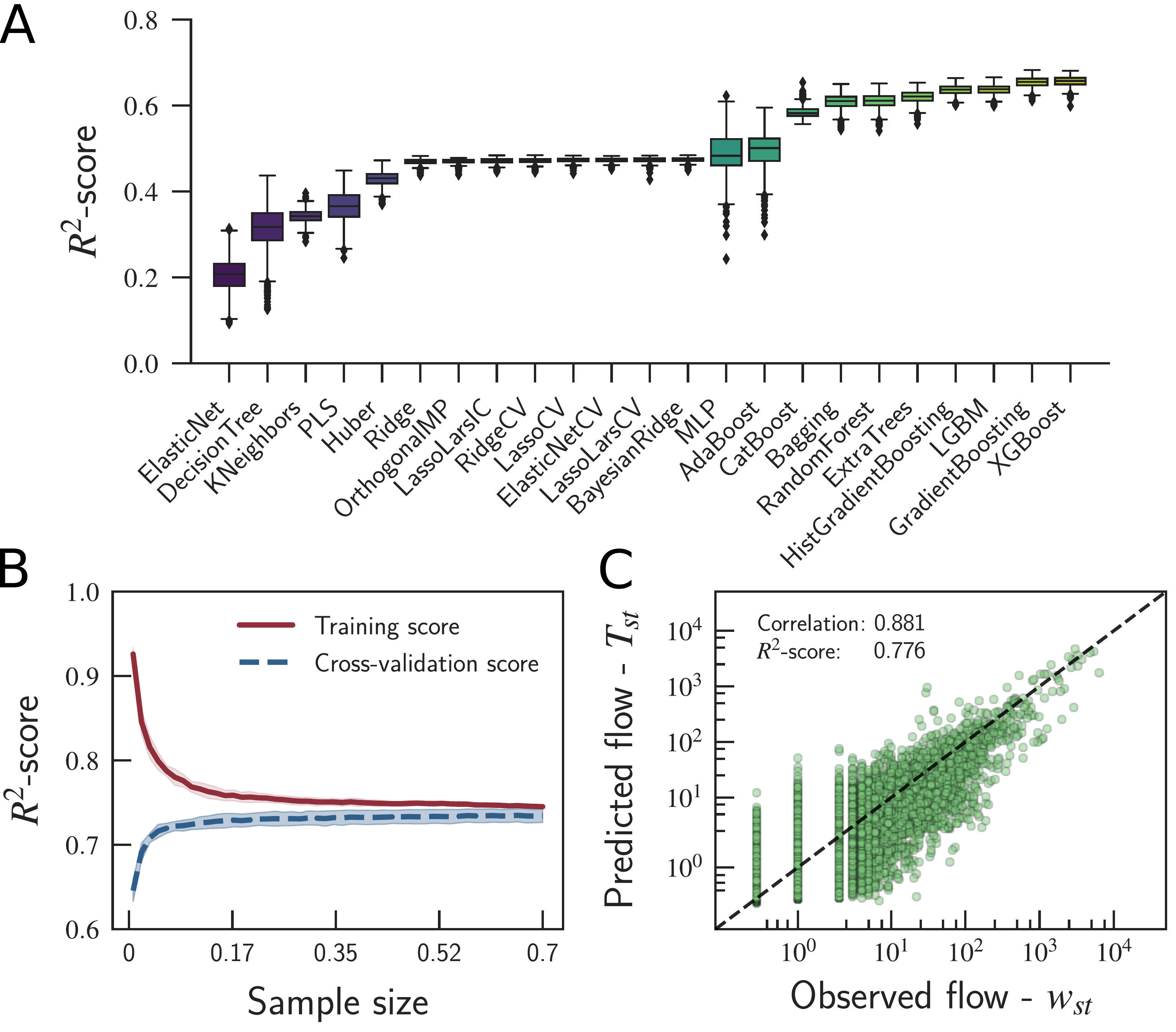}
    \caption{Performance of the regression algorithms in reconstructing the weighted projection of the commuters network; see Supplementary Table~2 for a list of the regressors' acronyms. {\bf (A)} $R^2$-score of the regressors' performance in quantifying the number of people ({\it i.e.}, $Weight_{st}$) commuting from city $s$ to city $t$. {\bf (B)} XGBoost learning curves varying the sample size. {\bf (C)} Predictions of $Weight_{st}$ made by the XGBoost regressor, after hyperparameter tuning, compared with the actual flow $w_{st}$. Specifically, our model as able to achieve an $R^2$-score of $77.6\%$ and a Pearson correlation coefficient of $0.881$.}%
    \label{fig:4}
\end{figure}

We used bootstrapping sampling to assess the variance of different regressors in terms of the $R^2$-score. From the 44 models, 21 were not able to provide all values (including outliers) of $R^2$-score higher than 0 (see Supplementary Table~2). Fig.~\ref{fig:4}A shows the results obtained from the 23 models that passed the first test, all of which were tested with default hyperparameter values from the {scikit-learn}~\cite{scikit-learn} and {XGBoost}~\cite{Chen2016:XGBoost} libraries. The models are presented in ascending order according to the value of their $R^2$-score. The results reveal that the three best algorithms are XGBoost, Gradient Boosting, and LGBM, and the three worst are Elastic-Net, Decision Tree, and K-Nearest Neighbors. The average $R^2$-score in the test varies from 20.6\% in the worst case ({\it i.e.}, Elastic-Net) to 65.6\% in the best case ({\it i.e.}, XGBoost), see Supplementary Table~4.

From now on, our focus will be on the best regression model, which was induced by XGBoost with an average $R^2$-score of 65.6\%, with no significant variance after a thousand predictions using different random samples of the training set. The further reason to choose the XGBoost as the best algorithm is that it is fast, as it provides a parallel tree boosting that solves problems faster than its competitors and has support to solve problems on high-performance computing environments. Thus, we evaluated the learning curve of the XGBoost regressor (see Fig.~\ref{fig:4}B), assessing the model predictability by varying the size of the training set, and we found that 100\% of the training data is required to train the regressor, yielding an $R^2$-score of 73.4\%. One can see that the regressors' predictions are more challenging than those of the binary classification algorithms, requiring more data during the training phase and yielding more errors.


Finally, we used the 30\% data reserved for testing the model. The results indicated that the regressor was able to predict $73.1\%$ of the variance observed in the data. After the hyperparameter tuning (see Methods), the XGBoost regressor described $77.6\%$ of the variance. Although the gains seem to be small, our model is intended to be used on data sets of thousands of cities, which answer for millions of possibilities of commuters flowing between pairs of cities. In such a case, 4.5\% improvement means weights closer to the real ones among the actual links. Fig.~\ref{fig:5}C compares the predictions about the number of commuters ($Weight_{st}$) with the actual values ($w_{st}$) observed in the data. Notice that, our model can recover the weighted network structure and predict the number of commuters with much higher precision than the gravitation and radiation models. In our case, even considering links with $w_{st}=0$, our predictions are much more accurate than those from the gravitation and radiation models (see Fig~\ref{fig:2}), which were induced using only existing flows.

\subsection*{Interpreting machine learning models using SHapley Additive exPlanations (SHAP)}

In contrast with the gravitation and radiation models, the proposed models can accurately reproduce the structure of the network, predicting whether a link between a given pair of cities exists or not (XGBoost classification) and accurately estimate the number of commuters between two cities (XGBoost regressor), reconstructing the weighted structure of the commuters network. The price of having a more complex model is that it becomes harder to interpret the relationship between the independent variables and our predictor. However, if we could understand how the algorithm makes decisions, the decision process would be a valuable resource to learn about the equations that describe our systems. Thus, instead of assuming mathematical relationships and trying to fit the model to the data, we could look into the results and try to figure out what the data tell us about the model that describes our system and the relationships between the commuter's flow and our features.

To turn our black-box algorithm in a more interpretative model, we calculated the features' importance and the impact of the features on individual predictions using the SHapley Additive exPlanations (SHAP) values~\cite{Lundberg2018:ExplainablePredictions,Lundberg2018:ConsistentAttribution,Lundberg2017:SHAP}. The SHAP metric is based on the Shapley values introduced by Lloyd Shapley in 1953 in the context of game theory~\cite{Shapley1953:PersonGames}; it helps one to understand how the model decides to make a prediction and which features contribute to improving the accuracy of the model. The course of action of SHAP is to calculate the importance of a feature by comparing what the model predicts with and without the feature. Notice that the order in which we add new features to the model can affect its predictions. Thus, we have to permute over all the possible feature orderings to fully capture the impact of a feature on the model.

\begin{figure}[!b]
    \centering
    \includegraphics[width=0.95\linewidth]{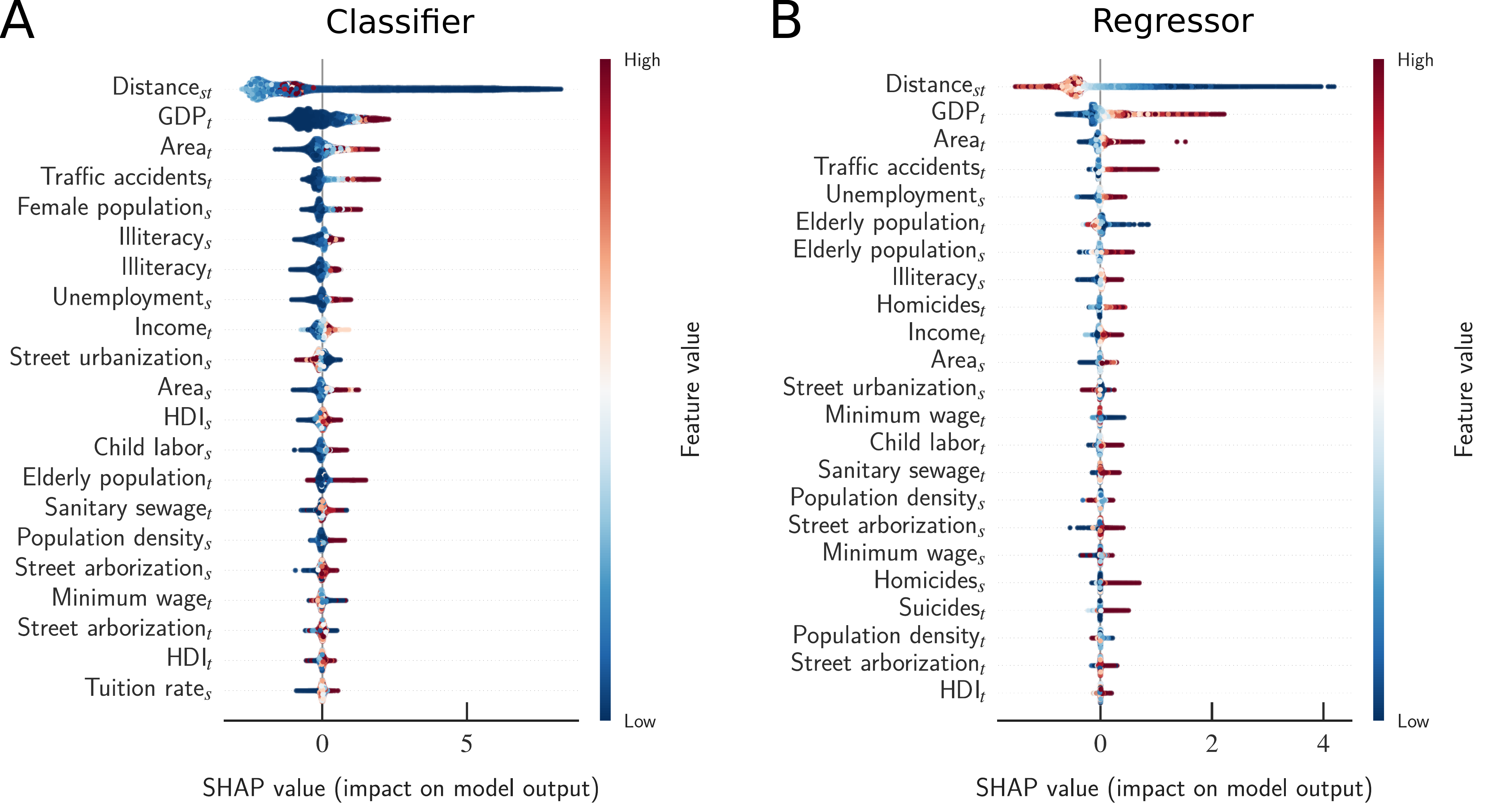}
    \caption{Interpreting the relationship between flow and features. Analysis of feature importance in the XGBoost {\bf (A)} classifier and {\bf (B)} regressor using SHAP values. The features are ranked by importance in descending order based on the sum of the SHAP values over all the samples. The violin-shaped plots show the distribution of the impacts of each feature on every point of the model output. The colors represent the SHAP value, varying from low values (blueish colors) to high values (reddish colors).}%
    \label{fig:5}
\end{figure}

In Fig.~\ref{fig:5}A, we show the features considered important for the XGBoost classifier, as well as the distribution of the impacts of each feature on the model output. While high values of distance $Distance_{st}$ (also known as $r_{st}$) are not good predictive features for the presence of links (values smaller than 0), high {\it GDP}$_{t}$ values increase the predictive accuracy of the model (values larger than 0). High rates of the elderly population in the target city also strongly affect the model accuracy.
Finally, the total urban area or traffic accidents are more important than the population density.

In Fig.~\ref{fig:5}B, we show the features' importance for the XGBoost regressor, as well as the distribution of the impacts of each feature on the model output. In this case, high values of distance $Distance_{st}$ have a low impact on the predictive performance of the model (values smaller than 0), whereas high {\it GDP}$_{t}$ increases the predictive performance of the model (values larger than 0). Higher values of the elderly population in the target city decrease the performance of the algorithm. Again, we verify that several urban indicators have a role in defining the flow of people from one city to others and that urban indicators, such like GDP, area, and traffic accidents, are more useful to improve the predictive performance of the model than population size.

\section*{Discussion}

Predicting decisions related to individual choices, such as housing and working places, is a difficult task. These decisions are tied to many variables that are often based on personal reasons and cannot be easily measured. However, the analysis of SHAP values can help us to understand how urban indicators and distance can influence this decision process and what makes people commute from one area to another to work. The SHAP values point four features that are mutually important in both classification and regression models, {\it i.e.}, Distance, GDP, Area, Traffic accidents in the target city, followed by other less important features.

{\it Distance} is the most important feature in predicting the existence of a link between a pair of cities and quantifying the number of people commuting from one city to the other. It is traditionally included in migration models. Distant cities make the journey to work unfeasible, as workers must return home by the end of the day, limiting the commuting to a certain distance. However, because of other factors, cities close to each other may have an insignificant flow, either because they offer similar opportunities or because the infrastructure and transport limit the commuting.

{\it GDP} (Gross Domestic Product) is the second most prominent feature. The cities with the highest GDP have the highest demand for commuters and the best job opportunities. As a consequence, they attract more commuters. Looking at Fig.~\ref{fig:5}, it is possible to observe that the higher the GDP value, the higher its contribution to the predictive performance of the XGBoost induced models. Another possible explanation is that these cities are more productive because they attract more workers and, therefore, they have higher GDP values, as suggested by Keuschnigg {\it et al.}~\cite{Keuschnigg2019:UrbanScaling} for long-term migration. 

{\it Area} is related to the physical size of a city, and it is the third most influential variable. Larger cities, like metropolis and megalopolis, usually are hubs of industries and enterprises and, consequently, tend to offer more job opportunities. This variable fails to affect the flow prediction when cities have a wide territorial extension but are used for other purposes ({\it e.g.}, natural reserves, forest, and huge urban parks) rather than urban expansion. Our data set reveals just a few records of cities occupying a large area with no significant number of commuters. These are the cases of cities with SHAP values close to 0, as we can observe in Fig.~\ref{fig:5}B.

{\it Traffic accident} is the fourth most important variable. This indicator was selected by the model because it has a high correlation with a higher flow of people and automobile vehicles (see Fig.~\ref{fig:1}C), which is also a trait related to the economics of the city, as is the case of GDP. Recall that, although the features we used are correlated, the feature selection process chose to keep the ones that it considers to be important, removing the other ones without negatively impact the predictive performance of the model (classifier and regressor).

The following features are ordered differently, depending on the predictive task, whether it is classification or regression. In the classification task, the classifier does not require much to predict a fair number of links correctly. For example, using just the previously discussed selected features on the training set, the classifier presented an accuracy close to 87.6\%. The other features improved the model accuracy by 1.6\%, also on the training set. Among all variables, the {\it unemployment} indicator of the source city ({\it i.e.}, where people live), seems to have a significant impact on the existence of flow between two cities. High rates of unemployment seem to make commuters leave the city where they live ({\it i.e.}, repel) to work in nearby cities with better job opportunities ({\it i.e.}, absorb).

Differently from the classification task, in the regression task, the algorithm demands more information to estimate the value of flow close to the actual value. Using only the four most important features, the model describes 70.6\% of the variance of the data on the training set. The remaining features are responsible for improving the regressor's predictive performance by 2.8\% on the training set. The {\it elderly population} indicator has an interesting behavior: the larger the number of elderly people in the home city and the smaller their number in the city of work, the higher the flow of commuters between the two.

The remaining urban indicators showed little relevance in predicting the flow of people, despite having a significant impact on the predictions of the classification and regression tasks. This set of features together provides a better prediction and highlights the importance of taking into account a more comprehensive set of indicators, in addition to the typical metrics used in migration models, such as population and distance. Further, such models could be enhanced by adding other indicators that were not explored in our data set, which could potentially improve the predictions of links and flow of commuters.

Finally, we believe that our approach is not only to be used on the task of commuters network reconstruction but also to complex networks of different domains where node's attributes can give us information about the existence of links connecting them. For instance, this methodology could be applied in economic trade networks~\cite{a2018unfolding,alves2019nested}, where the amount of money exchanged could be predicted in terms of indicators as GDP, unemployment, interest rate, production, corporate profits, and other macroeconomic variables; in social networks\cite{adamic2003friends}, where friendship connections could be unveiled by individual node's preferences, such as music or sport preferences, language or individual socioeconomic status and education; in metabolic networks~\cite{guimera2005functional}, where biochemical reactions (links) could be predicted in terms of metabolites chemical, physical, and biological properties. Thus, further insights from these networks could be obtained by exploring the SHAP values to identify decisive indicators to reconstruct each network topology.

\section*{Methods}

\paragraph{Data set.}

We collected data about the number of people commuting from a home city to another city to work (pendular migration) in the year of 2010, considering all the $5,565$ Brazilian cities as well as 22 yearly-updated urban indicators. This data is provided and maintained by the Brazilian Institute of Geography and Statistics (IBGE)~\cite{IBGE}. The data was modeled as a complex network represented as a directed graph $G = \left\{V, E\right\}$ composed of $5,565$ vertices and $55,247$ edges with non-zero weights. A vertex $s \in V$ is considered to be a city, and an edge $e \in E$ is an ordered pair $e = \left<s, t\right>$ representing the flow from the source city $s \in V$ to the target city $t \in V$. The existence of commuters from city $s$ to city $t$ is noted as $Class_{st}=1$, and $Class_{st}=0$ indicates the opposite; the flow $Weight_{st}$ is an integer representing the number of commuters that move daily from city $s$ to city $t$. The urban indicators shape an initial 45-value feature vector. This vector is composed of urban indicators (a single value per indicator) from the source and target cities (22 values for each city) and their distance (in kilometers).

\paragraph{Data split.}

The data set composed by the number of commuters (target feature) and the $2 \times 22$ urban indicators plus the distance between pairs of cities (predictive features) was split in a $70/30$ ratio preserving the ratios of existing and non-existing links in both sets. Thus, $70\%$ was used as a training set, and the other $30\%$ was used as a test set to evaluate the predictive performance, generalization, and universality of the model induced after all training and selection steps. To assure that the model could learn all inherent nuances of our data ({\it e.g.}, different intercity road systems), we also stratified the data by the Federal Brazilian States to feed the model with a proportional amount of information about every state. This also saves computational time to train the models by eliminating most of the non-existent links across states.

\paragraph{Class balancing.}

Because imbalanced data sets favor the majority class, we balanced the classes of the resulting data set keeping $50\%$ of the data labeled with existing links, $w_{st}\neq 0$, and the other with non-existing links $w_{st}=0$, resulting in $110,494$ pairs of cities (edges). For each existing connection between a city from state A and another from state B, our data set has a non-existing link between cities (chosen at random) from these states. This process was carried out for both tasks, classification, and regression. In the classification, the labels are $Class_{st}=1$ if there is a non-zero flow and $Class_{st}=0$ otherwise, whereas, in the regression task, the data was labeled with the number of commuters in the flow from one city to the other.

\paragraph{Training set stratification.}

We used a label-based stratified $k$-fold with 5 folds for the classification tasks (categorical data) and a regular $k$-fold with $5$ folds for the regression tasks (number of commuters).

\paragraph{Model selection.}

Using the training set, which represents $70\%$ of the whole data set and is composed of $77,345$ edges, we tested 78 algorithms of machine learning, including 34 classifiers and 44 regressors. In this step, we used the validation subset of the training data to select the best classifier to predict the existence of a non-zero flow (link) between cities and the best regressor to predict the number of commuters flowing between them (links' weight). This task was carried out using Bootstrapping, which is a statistical resampling method that consists of sampling with replacement to define the upper and lower bound of a statistical evaluation metric (accuracy and $R^2$-score)~\cite{Efron1992:Bootstrap}. We used this method to evaluate different models over random samples of the data set, without compromising the significance of the results. From all classifiers and regressors, we only used those that could yield a predictive performance without needing early hyperparameter tuning. Almost one-third of all algorithms did not meet such constraints, leaving 27 classifiers and 23 regressors for the experiments. The remaining algorithms went through a bootstrapping sampling by training each one with tiny random samples (1\% of the data chosen with replacement, {\it i.e.}, $7,734$ edges) of the training set (70\% of the whole data set, {\it i.e.}, $77,345$ edges) to access the variance of predictions over a thousand iterations. Through the bootstrapping sampling results, we defined the upper and lower bound of the scoring metric (accuracy and $R^2$-score). Using these metrics as criteria, we select the classifier and regressor with the highest median, lowest variance, and fewer outliers. It is important to note that bootstrap sampling uses random samples of 1\% of the training set to increase the randomness within each test sample so to capture all nuances related to the variance and outliers among thousands of predictions.

\paragraph{eXtreme Gradient Boosting (XGBoost).}

XGBoost is a machine learning library that provides a set of techniques to deal with and model data through both classifiers and regressors. Such a library was designed to be highly efficient, flexible, and portable. All the model-inferring algorithms were implemented under the Gradient Boosting framework providing further support to parallel and distributed processing~\cite{Chen2016:XGBoost}. Gradient Boosting, on the other hand, renders a predictive model through tree ensembles, which are submitted to optimization through an arbitrary differentiable loss function~\cite{Friedman2001:GradientBoosting}.

\paragraph{Learning curves.}

We used analysis of learning curves to assess the prediction capability of each model ({\it i.e.}, XGBoost classifier, and regressor) when increasing the size of the training data in increments of 1\% up to 70\%. The right amount of data to train each model is the one with the best trade-off between bias and variance, that is, where the training score curve and the cross-validation curve have the lowest deviation.

\paragraph{Feature selection.}

We performed a threshold-based feature selection to decrease the complexity of the induced model by removing features from the least important to the most important one up to a dynamically-defined importance-threshold. Such a process provides a subtler interpretation of the features' impact on the model's output and is carried out together with the XGBoost algorithm, which provides an importance score to each feature during the training phase. The importance score is defined in the interval $[0, 1]$, where $0$ indicates a feature with minimum or no importance, and $1$ indicates the opposite. The importance score consists of counting the number of times each feature is used to split a node into two trees by the XGBoost algorithm during the training phase. The feature selection process consists of measuring the model's prediction score while progressively removing the features from the least important to the most important one. These iterations define a threshold that splits the interval into two, such that features with importance between 0 and the threshold are removed. The threshold is defined by iteratively increasing the value from $0$ up to $1$ until the model's prediction score using all features starts decreasing. It is important to note that the features removed are not relevant to the induced model because they do not add useful information to predict the target data. This comes from the fact that urban indicators are collinear to each other, which means each indicator might contain traces of other indicators within it. Bettencourt~\cite{Bettencourt2013:ScalingCities} already discussed such phenomena on urban indicators, however, there still no way to avoid or remove it completely from such type of data. The threshold-based feature selection works around this problem by removing non-relevant and collinear features, which absence impacts the model nor positively neither negatively. Therefore, the benefit of using feature selection, in this case, is its ability to reduce the model complexity by decreasing the number of dependable variables.

\paragraph{Hyperparameter tuning.}

For XGBoost classifier and regressor, we performed hyperparameter tuning; a brute-force technique that seeks for the set of hyperparameters within a pre-defined hyperparameters' subspace that optimizes a provided machine learning model. For the sake of experimentation, we tested 3 million possibilities out of a hyperparameters space that is exponentially larger. Such a fine-tuning is considerably time-consuming and can take several weeks of intensive computing for each model. On the other hand, the time is proportional to the number of tested hyperparameters, which is a fair trade-off, and that can be decreased with high-performance computing.

For our experiments, we found that the best hyperparameter values for the XGBoost classifier were: {\it (i)} learning rate $=0.05$; {\it (ii)} number of trees $=300$; {\it (iii)} maximum tree depth $=9$; {\it (iv)} fraction of observations used as random samples by each tree $=0.85$; {\it (v)} subsample ratio of features when constructing each tree $=0.8$; and, {\it (vi)} regularization term of the model's weights $=0.75$. In the case of the XGBoost regressor, the best hyperparameter values  were: {\it (i)} learning rate $=0.05$; {\it (ii)} number of trees $=900$; {\it (iii)} maximum tree depth $=7$; {\it (iv)} fraction of observations used as random samples by each tree $=0.85$; {\it (v)} regularization term of the model's weights $=1.25$; {\it (vi)} minimum loss reduction required to split a node into two trees $=0.25$; and, {\it (vii)} minimum weight needed in a child node $=2$.

\paragraph{Model evaluation.}

After all the training and selection steps, we used the testing set to compare the non-optimized and optimized models, reporting the results in a confusion matrix for the classifier and a scatter plot for the regressor. The classifiers were evaluated through the balanced accuracy score ({\it i.e.}, the ratio between the correct classified instances and all instances, normalized by the number of elements per label) and the regressors through the $R^2$-score ({\it i.e.}, the variance between the predicted and observed values). 

\paragraph{SHAP values and feature importance.}

Given a specific prediction $f(x)$, we calculate the importance of a feature by comparing what a model predicts with and without the feature. Because the order in which we add new features to the model can affect its predictions, we permuted over all possible feature ordering. Mathematically, the Shapley value for a particular feature $i$ (out of $M$ total features), given a prediction $x$ is:
\begin{equation}
    \centering
    \phi_i(f,x)= \sum\limits_{S \subseteq M \setminus \{i\}} \frac{|S|!(M - |S| -1)!}{M!} \left[f_x(S \cup \{i\}) - f_x(S)\right],
    \label{eq:shapley}
\end{equation}
\noindent where $S$ is the set of features used to induce the model, and $f_x$ is the prediction considering the indicated set of features~\cite{Lundberg2018:ExplainablePredictions}. In practice, this is too difficult to be calculated because there are far too many possible combinations. Instead, we used the SHAP library to calculate $\phi_i$, which is optimized to take advantage of different model’s structures~\cite{Lundberg2017:SHAP}. Thus, the rank of feature importance is given by the sum of SHAP value magnitudes $\phi_i$ over all samples. This procedure allows a more in-depth interpretation of the impact of the features on the prediction of individual data, turning the black-box algorithm in a more interpretable model.

\section*{Acknowledgments}
The authors would like to thank Coordena\c{c}\~ao de Aperfei\c{c}oamento de Pessoal de N\'ivel Superior - Brazil (CAPES) - Finance Code 001; Funda\c{c}\~o de Amparo \`a Pesquisa do Estado de S\~ao Paulo (FAPESP), through grants 2013/07375-0, 2014/25337-0, 2016/16987-7, 2016/17078-0, 2017/08376-0, and 2019/04461-9; Conselho Nacional de Desenvolvimento Cient\'ifico e Tecnol\'ogico (CNPq) through grants 303694/2015-7, 404870/2016-3, 167967/2017-7, and 305580/2017-5; and Intel for their financial support. They also would like to thank Daniel M. Lima, Lucas C. Scabora, and Willian D. Oliveira for their help with the Edge Bundling algorithm.

\section*{Author contributions statement}

GS and LGAA conceived the experiments. GS conducted the experiments. GS and LGAA analyzed the results. GS, ACC, JFR, and LGAA validated the results. GS and LGAA wrote the original draft. All authors reviewed and approved the manuscript.

\section*{Competing Interests} 
The authors declare no competing interests.


\end{document}